\documentclass[%
superscriptaddress,
bibnotes,
 amsmath,amssymb,
 aps,
 prl, 
longbibliography,
eprint,
floatfix,
reprint
]{revtex4-2}

\usepackage{xcolor}
\definecolor{darkblue}{RGB}{0,0,150}
\definecolor{nightblue}{RGB}{0,0,100}

\usepackage{mathrsfs,dsfont,mathtools,braket,MnSymbol,graphicx}

\newcommand{\refsub}[2]{\hyperref[#1]{\ref*{#1}#2}}

\usepackage{bm}
\usepackage[
colorlinks,
citecolor=darkblue,
linkcolor=darkblue,
urlcolor=nightblue]{hyperref}

\usepackage[english]{babel}
\usepackage[babel,kerning=true,spacing=true]{microtype}

\definecolor{DarkRed}{RGB}{100,0,0}
\definecolor{LightGreen}{RGB}{000,50,0}

\def\({\left(}
\def\){\right)}

\def\inf{\infty}

\newcommand{\beq}{\begin{equation}}
\newcommand{\eeq}{\end{equation}}
\newcommand{\bal}{\begin{aligned}}
\newcommand{\eal}{\end{aligned}}

\newcommand{\bfr}{{\bf r}}

\begin{document}
\title{
Superdielectrics: Disorder-induced perfect screening in insulators
}

\author{Ilia Komissarov}
\email{i.komissarov@columbia.edu}
\affiliation{Department of Physics, Columbia University, New York, NY 10027, USA}
\author{Tobias Holder}
\email{tobiasholder@tauex.tau.ac.il}
\affiliation{School of Physics and Astronomy, Tel Aviv University, Tel Aviv, Israel}
\author{Raquel Queiroz}
\email{raquel.queiroz@columbia.edu}
\affiliation{Department of Physics, Columbia University, New York, NY 10027, USA}
\affiliation{Center for Computational Quantum Physics, Flatiron Institute, New York, New York 10010, USA}

\date{\today}

\begin{abstract}
We study the relationship between the quantities that encode the insulating properties of matter: the ground-state quantum metric, the average localization length, and the electric susceptibility. By examining the one-dimensional Anderson insulator model and the Su-Schrieffer-Heeger chain with chiral disorder, we demonstrate that the former two measures are proportional in one-dimensional systems near criticality, and both are determined by the properties of the hybridized localized states around the Fermi energy. We employ these insights to demonstrate that the behavior of the electric susceptibility is drastically different in the bond-disordered SSH chain, with the possibility that it may diverge even when the localization length and the quantum metric remain finite. This divergence, caused by the proliferation of impurity resonances at a particular energy, leads to a novel regime that exhibits mixed characteristics of metals and insulators. We term this regime \emph{superdielectric}: an insulating state characterized by a finite quantum metric and divergent static electric susceptibility, which implies perfect screening in the absence of the dc conductivity. We demonstrate that the superdielectric phase also emerges in higher-dimensional materials, such as graphene with vacancies and Kekul\'e bond distortion.
\end{abstract}

\maketitle

\paragraph{Introduction.---} The ground state quantum geometric tensor $\mathcal{Q}^{\mu \nu}$ is a fundamental quantity that distinguishes metals and insulators \cite{resta_geometric}: its real and symmetric part, $g^{\mu\mu}=\mathrm{Re}\,\mathcal{Q}^{\mu\mu}$, is the quantum metric, which can be obtained from the longitudinal ac conductivity $\sigma^{\mu \mu}(\omega)$ via the Souza-Wilkens-Martin sum rule \cite{swm, rs}
\begin{equation}
g^{\mu\mu}\equiv \frac{1}{\pi} \int_0^\infty d \omega \frac{\sigma^{\mu\mu}(\omega)}{\omega}\, . 
\end{equation}  
By this definition, $g \equiv g^{xx}$ remains finite in insulators, when dc conductivity vanishes $\sigma^{xx}(0) = 0$, while finite dc conductivity in metals implies that $g$ diverges \cite{kohn}. The quantum metric $g$ has recently attracted growing interest due to its crucial role in electronic transport and correlated phenomena \cite{qg1, qg2,qg3,qg4,Jiang2025, nagaosa}. In crystalline insulators, it is well understood that $g$ arises from ground state dipole fluctuations of the electronic wavefunction within the unit cell, with the characteristic scale $\ell_{g}= \sqrt{n^{-1} g} \sim a$, reflecting the average spatial spread of electrons in the ground state, where where $n$ is the electron density and $a$ is the lattice constant. The quantity $\ell_g^2$ is also known as the localization tensor \cite{kudinov, swm, resta_geometric}.

\begin{figure}[t!tbp]
    \centering
    \includegraphics[width=0.45\textwidth]{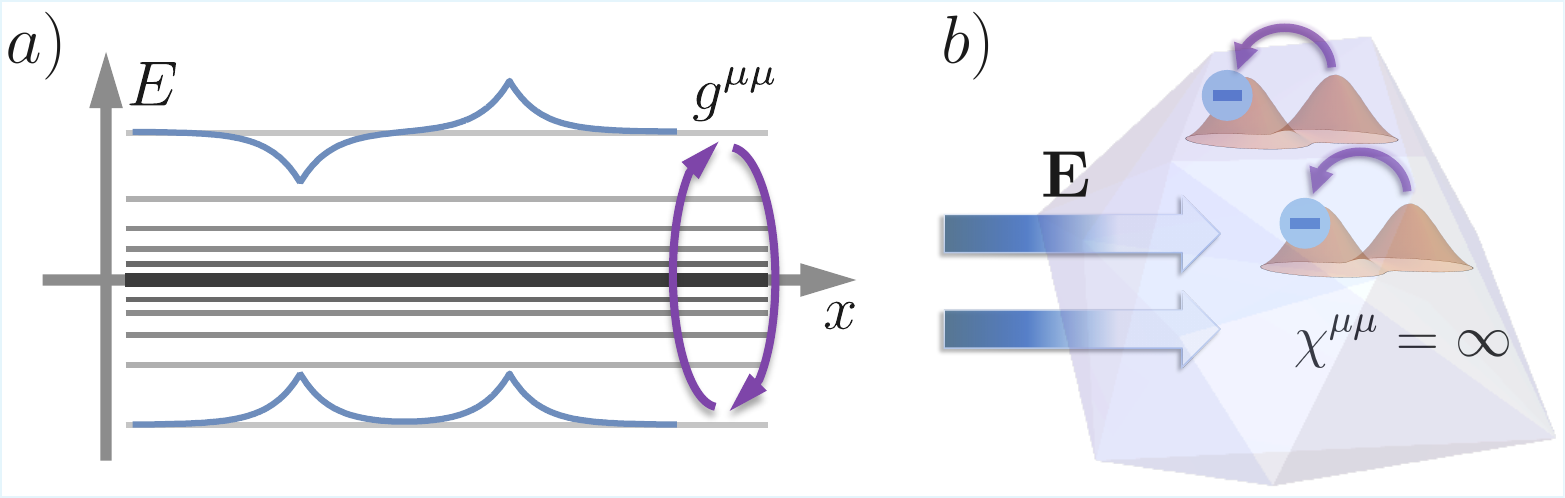}
    \caption{$a)$ In chiral disordered insulators, such as the SSH chain with bond disorder, impurity-induced zero modes accumulate at the same energy and hybridize pair-wise, forming double-peak states. Away from criticality, optical transitions between such states give rise to the finite $g^{\mu\mu}$. In 1D, the quantum metric in this system is proportional to the average localization length $g \sim \xi_{\rm av}$. $b)$ Despite the quantum metric being finite (and therefore, the dc conductivity vanishing), the electric susceptibility $\chi^{\mu \mu}$ in chiral disordered insulators may diverge, indicating perfect metal-like screening of externally applied static electric field $\mathbf E$. This mechanism is facilitated by electron hopping between far-away localized impurity states.}
    \label{fig1}
\end{figure}

Despite the abundance of applications of the quantum metric, e.g., in bounds for the dielectric constant \cite{komi} or the topological gap \cite{gapbd}, the understanding of its origin and scaling in disordered systems remains lacking, and the studies addressing it \cite{qmdis1, qmdis2, qmdis3}  remain sparse. In this work, we explore the behavior of the quantum metric in one-dimensional disordered systems, showing that $g$ is dominated by the optical transitions between resonant hybridized impurity states, and its value is governed by the average localization length $\xi_{\rm av}$, which sets the scale for the spatial decay of the disorder-averaged conductivity \cite{kramer, evers_anderson_2008}. Specifically, we demonstrate that in both the 1D Anderson model, and the Su-Schrieffer-Heeger (SSH) chain \cite{ssh} with bond disorder, optical transitions between localized impurity states lead to the scaling $g \sim \xi_{\rm av}$ near criticality, or $\ell_g \sim \sqrt{a \cdot \xi_{\rm av}}$, i.e., the average size of the occupied electronic eigenstate is determined by the geometric average of $\xi_{\rm av}$ and the lattice constant $a$. We verify this scaling both numerically and analytically, using the resummation of the impurity-scattering diagrams introduced in \cite{berez} in the Anderson model and the zero-mode hybridization arguments outlined in \cite{komi2} in the bond-disordered SSH chain. 

Crucially, in the SSH chain with bond randomness, the typical $\xi_{\rm typ}$ and average $\xi_{\rm av}$ localization lengths behave differently since only the average is sensitive to the effects of the \emph{rare topological zero mode hybridization events}, which induce critical behavior in this system \cite{komi2} appearing due to the accumulation of modes at zero energy, not present in the Anderson model \cite{evers_anderson_2008}. We demonstrate that the quantum metric naturally captures the effect of such rare events, providing an alternative means to extract $\xi_{\rm av}$ from the known eigenvalues of the Hamiltonian, thereby avoiding the arduous computation of the disorder-averaged conductance. The latter quantity is challenging to access numerically due to its non-self-averaging character. The quantum metric, on the other hand, is not purely a Fermi surface property, and the effect of the disorder fluctuations on it is much more tamed, which makes its computation more numerically stable. 

\begin{table}[t!]
\begin{center}
\begin{tabular}{|c|c|c|c|}
\hline
\multicolumn{4}{|c|}{Critical scaling of localization measures in 1D} \\
\hline
model & Anderson \eqref{andham} & SSH \eqref{sshham} ($E_{\rm F}=0$) & SSH \eqref{sshham} ($\delta=0$) \\
\hline
$\xi_{\rm typ}$ & $\sim W^{-2}$ \cite{economou2006green} & $\sim \delta^{-1}$~\cite{Note1} & $\sim \log E_{\rm F}$~\cite{Note1} \\
\hline
$\xi_{\rm av}$ & $4\, \xi_{\rm typ}$ \cite{evers_anderson_2008} & $\sim \delta^{-2}$ \cite{evers_anderson_2008} & $\sim \log^2 E_{\rm F}$~\cite{Note1} \\
\hline
$g$ & $c_g\, \xi_{\rm typ}$~\cite{Note1} & $\sim \delta^{-2}$ \eqref{gcrit} & $\sim \log^2 E_{\rm F}$~\cite{Note1} \\
\hline
$\chi$ & $c_{\chi} \nu({E_{\rm F}}) \xi_{\rm typ}^2$ \cite{ feigelman_dielectric_2018} & --- & $\sim E_{\rm F}^{-1} \log E_{\rm F}$~\cite{Note1} \\
\hline
\end{tabular}
\end{center}
\caption{Critical scaling of the typical $\xi_{\rm typ}$ and the average $\xi_{\rm av}$ localization lengths, the quantum metric $g$, and the electric susceptibility $\chi$ in the studied one-dimensional models with disorder: the Anderson model \eqref{andham}, and the SSH chain with bond disorder \eqref{sshham}. In all studied cases, the critical scaling $g \sim \xi_{\rm av}$ holds. The proportionality constant in the expression for the susceptibility is $c_\chi = 4 \zeta(3) \simeq 4.81$, and the one for the quantum metric is $c_g \simeq 0.1289$, as obtained diagrammatically within the continuum model with the Gaussian noise disorder, and $c_g \simeq 0.14$ is the numerical estimate for the full Anderson model~\cite{Note1}.}
\label{tab1}
\end{table}

We also analyze the electric susceptibility $\chi \equiv \chi^{xx}$, which, like $g$ and $\xi_{\rm av}$, probes ground‑state localization. Typically, $\chi$ is finite in insulators and diverges in metals \cite{cat1}, as for example observed in doped silicon \cite{cat2}. It is, however, much more sensitive to the states around the Fermi energy. In systems with chiral disorder, a buildup of impurity states at a single energy \cite{tr1, tr2} causes $\chi$ to diverge while $g$ stays finite (Fig.~\ref{fig1}). This behavior arises from rare, large hybridized zero‑mode pairs with exponentially small level splittings appearing with a probability that scales exponentially with their size \cite{komi2}—i.e., due to the Griffiths effects \cite{vojta}. We term the insulating materials featuring metal‑like static screening with vanishing dc conductivity \textit{superdielectrics}. We show that the examples of such systems include the bond‑disordered SSH chain as well as the Kekul\'e distorted graphene with vacancies, both of which show a singular density of states \cite{gadevac, grres}, forcing the average optical gap to vanish $\overline{\Delta}=2g/\chi \to 0$.

This letter is organized as follows: we begin by demonstrating the proportionality between the quantum metric and the average localization length at criticality in the Anderson chain. Despite the absence of a spectral gap in this model ($\Delta = 0$), we argue that the average optical gap $\overline{\Delta}$ remains nonzero, leading to finite $\chi$. We then show that a similar critical scaling relation, $g \sim \xi_{\rm av}$, holds in the SSH chain with chiral disorder. We then demonstrate that, unlike the Anderson insulators, both the chiral-disordered SSH chain and vacancy-doped Kekul\'e graphene exhibit a superdielectric phase due to the presence of rare and extended impurity resonances. 

\paragraph{Relation between $g$, $\xi_{\rm av}$, and $\chi$ in the Anderson chain.---} As a warm-up, we demonstrate the proportionality of the quantum metric and the average localization length in the 1D Anderson model described by the Hamiltonian
\begin{equation}
\hat H_{\rm Anderson} = \sum_{i} t (\ket{i}\bra{i+1} + {\rm h.c.}) + \sum_{i} \varepsilon_i \ket{i} \bra{i} \, ,
\label{andham}
\end{equation}
where the on-site potentials $\varepsilon_i$ are randomly distributed around zero with $\llangle \varepsilon_i^2 \rrangle \sim W^2$, where double brackets denote the disorder average. At any disorder strength $W$, all the states in this model are exponentially localized \cite{mottbook, erdos, berez, apricot}, and therefore the disorder-averaged quantum metric
\begin{equation}
g = \frac{1}{L} \sum_{E_m < E_{\rm F} < E_n} \llangle |\braket{n|\hat x|m}|^2 \rrangle
\label{g}
\end{equation}
is finite for any Fermi energy $E_{\rm F}$ \cite{resta_geometric}. In \eqref{g}, $\ket{n}$ are eigenstates with energy $E_n$, and $L$ is the length of the chain. Sufficiently close to criticality ($\xi_{\rm av} \gg 1$), we expect the dominant contribution to the quantum metric to arise from the optical transitions between the impurity states around the Fermi surface. Since the only relevant length scale that determines the average localization size of states near $E_{\rm F}$ is $\xi_{\rm av}$, based on the dimensional analysis, we expect $g \sim \xi_{\rm av}$. We confirm this scaling analytically using the low-energy continuum model corresponding to \eqref{andham} with the Gaussian noise disorder, which can be analyzed in the limit $\xi_{\rm typ} \gg 1$ using the diagrammatic resummation \cite{berez}, as shown in Supplementary Information~\footnote{See supplementary information for the list of notations, a summary of the diagrammatic resummation approach, the relation between the dynamical structure factor and the transport quantities, discussion of the zero mode resonances in the SSH chain, and details of transfer matrix calculations in the Kekul\'e-distorted graphene model.}. Utilizing this formalism, we obtain 
\begin{equation}
g \simeq 0.1289 \cdot \xi_{\rm typ} = 0.03223 \cdot \xi_{\rm av} \, ,
\label{gxi}
\end{equation}
where we accounted for the fact that the typical localization length $\xi_{\rm typ}$ is related to the average localization length in this model as $\xi_{\rm av} = 4 \xi_{\rm typ}$ \cite{evers_anderson_2008, apricot}. 

We test this relation numerically by calculating the quantum metric $g$, using \eqref{g} and the typical localization length $\xi_{\rm typ}$ for different disorder distributions and values of the Fermi energy $E_{\rm F}$ (see ~\cite{Note1} for the numerical data). The obtained for the full Anderson model \eqref{andham} values of the ratio $g/\xi_{\rm typ} \simeq 0.14$ agree approximately with the analytically predicted continuum model coefficient due to the crucial and universal role played by the optical transitions between the low-lying in energy spatially-extended Mott resonances~\cite{Note1}.
 
Note that the electric susceptibility in the Anderson model \eqref{andham} is given by $\chi \simeq 4.808 \cdot \nu(E_{\rm F}) \xi_{\rm typ}^2$ \cite{feigelman_dielectric_2018}. Therefore, even though this system lacks the spectral gap $\Delta$, the average optical gap $\overline \Delta = 2g/\chi \simeq (20 \cdot \nu \xi_{\rm typ})^{-1}$ remains finite away from criticality. This is in contrast to systems with chiral disorder, where the build-up of states at a particular energy causes $\overline \Delta$ to vanish even at a certain distance from criticality, as we show below.

\begin{figure}[t!]
    \centering
    \includegraphics[width=0.45\textwidth]{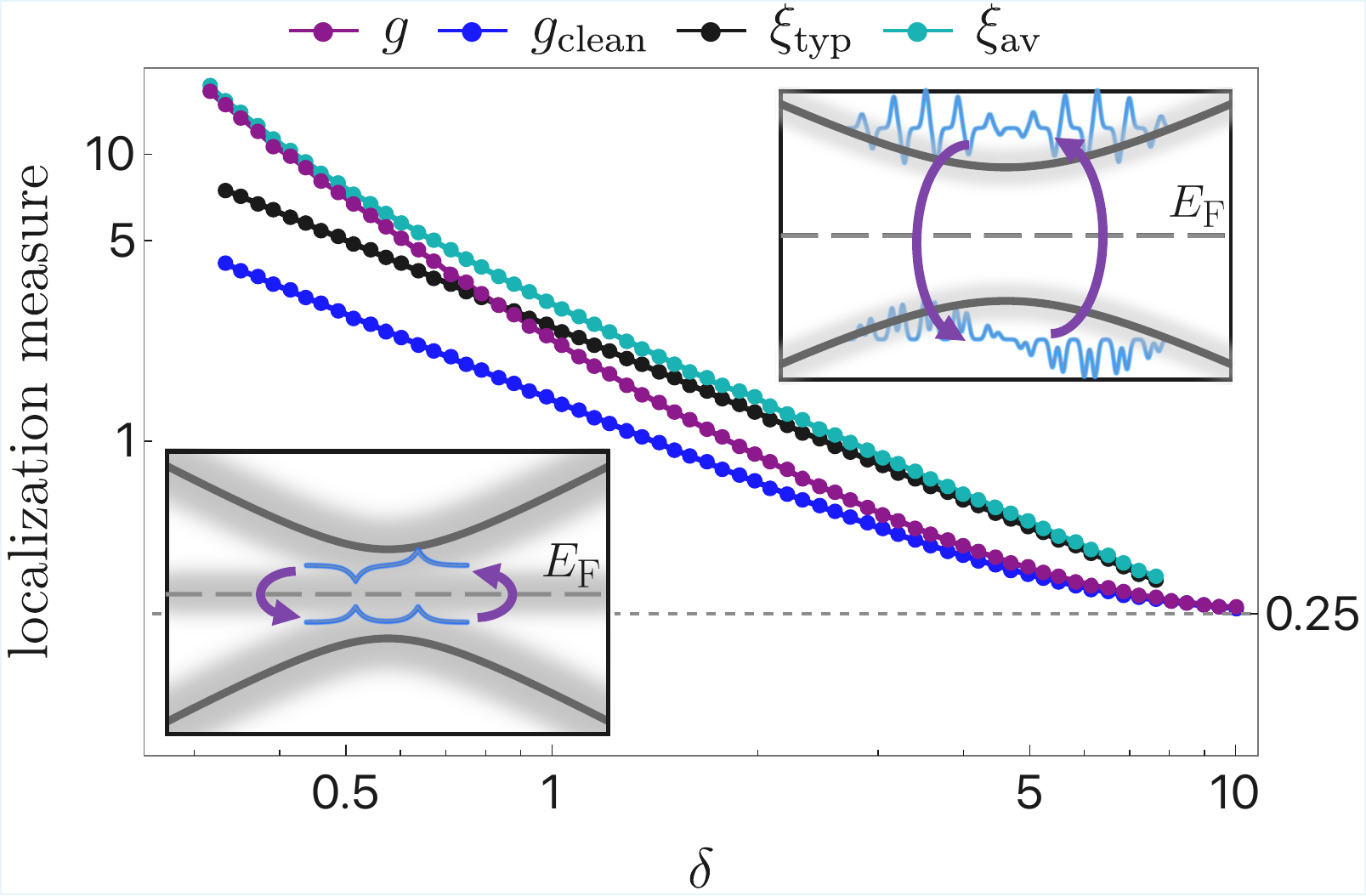}
    \caption{Different localization measures computed in the SSH chain with chiral box disorder of the strength $s^{2} = 1/5$ and without disorder ($s^2=0$, denoted with ``clean'') as functions of the hopping dimerization parameter $\delta$. At small $\delta$, by fitting we establish the scaling $g \sim\xi_{\rm av} \sim \delta^{-2}$ and $g_{\rm clean}\sim \xi_{\rm typ}\sim \delta^{-1}$. The enhanced scaling of $g \sim \xi_{\rm av}$ is due to the hybridized topological zero modes (see Eq.~\eqref{gcrit}) shown in the inset on the left. At high dimerization, the disorder is irrelevant, and we observe $g \simeq g_{\rm clean} \simeq 1/4$, which encodes the quadratic spread of the Wannier orbitals delocalized between the ends of almost isolated dimers. The interband transitions shown with the purple arrows in the inset on the right dominate the value of $g$ in this case.}
    \label{fig2}
\end{figure}

\paragraph{Relation between the quantum metric and the localization length in the SSH chain.---} In the following, we consider the SSH chain with hopping disorder described by the Hamiltonian \cite{meier, windingr}
\begin{align}
    \hat H_{\rm SSH} &= \sum_i (t' - \varepsilon_i')\ket{i,\rm B} \bra{i,\rm A}  
    \notag\\&\qquad + (t - \varepsilon_i) \ket{i+1,\rm B} \bra{i,\rm A} + {\rm h.c.}\, ,
    \label{sshham}
\end{align}
with $\varepsilon_i$ and $\varepsilon_i'$ randomly distributed with $\llangle \varepsilon_i \rrangle = \llangle \varepsilon_i' \rrangle = 0$, $\llangle \varepsilon_i^2 \rrangle \sim \llangle \varepsilon_i'^2 \rrangle \sim W^2$. When $|t| = |t'|$, this model hosts a single delocalized state at zero energy \cite{eggarter, theodorou_extended_1976}, while all the states are exponentially localized otherwise. We concentrate on the most physically interesting half-filled case ($E_{\rm F} = 0$) for which the system may enter criticality. It is convenient to introduce the quantities $s^2$ and $\delta$, which parameterize the disorder strength and the hopping dimerization:
\begin{equation}
s^2=\llangle \kappa^2 \rrangle - \llangle \kappa \rrangle^2,~~ \delta = \llangle \kappa \rrangle/s^2, ~~ 
    \kappa \equiv \log \left|\frac{t - \varepsilon_i}{t' - \varepsilon_i'}  \right| \, .
\end{equation}
In these variables, criticality is reached for $\delta = 0$ at any value of $s$.

\begin{figure}[t!]
    \centering
    \includegraphics[width=0.4\textwidth]{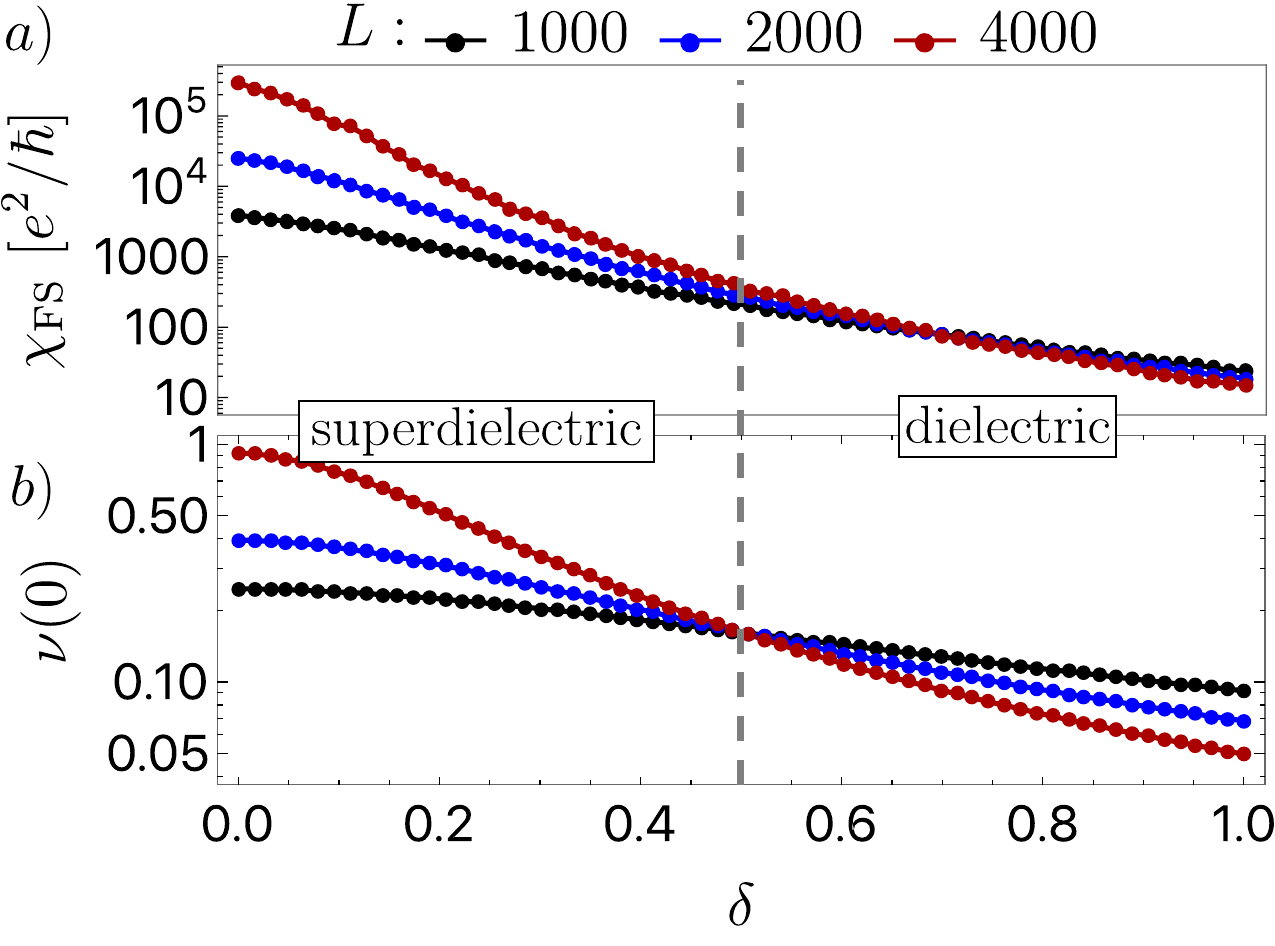}
    \caption{Behavior of the Fermi surface contribution to electric susceptibility $\chi_{\rm FS}$ ($a$) and the density of states at zero energy $\nu(0)$ ($b$) against the hopping dimerization strength $\delta$. At $0< \delta \leq 1/2$, the finite-size scaling is consistent with $\nu(0) = \chi_{\rm FS} = \infty$, indicating an insulator with divergent electric susceptibility, a superdielectric. For $\delta > 1/2$, we obtain $\nu(0) = 0$, and finite $\chi_{\rm FS}$ confirming the regular insulating dielectric behavior. Note that $\chi_{\rm FS} \sim \nu(0)$ diverge exponentially in volume at small $\delta$, as expected from the relation between the energy and the size of the zero-mode Mott resonances, which gives rise to the finite-size cutoff $E \sim e^{-L}$ \cite{komi2}. To estimate $\chi_{\rm FS}$, we employed the Fermi surface wavefunction autocorrelations~\cite{Note1}.}
    \label{fig3}
\end{figure}

Let us consider how the quantum metric $g$ behaves when the dimensionless disorder strength $s^2$ is fixed at a constant value while the dimerization parameter $\delta$ changes. We expect that for $\delta \gg 1$, the disorder is irrelevant, and the quantum metric can be approximated by its disorder-free value $g_{\rm clean} \equiv g(\delta, s^2 = 0)$, which saturates to $1/4$ in the dimerized limit corresponding to the quadratic spread of the Wannier orbitals \footnote{The Wannier function localized on an $i$-th isolated dimer can be expressed as $\ket{w_i} = \sqrt{2}^{-1} (\ket{i, \rm A} + \ket{i, \rm B})$, and the quadratic spread is found as $\braket{w_i|\hat x^2|w_i} - (\braket{w_i|\hat x|w_i})^2 = a^2/4=1/4$}. For small dimerization $\delta$ (the dirty limit), the contribution due to impurity states becomes important, and --- analogously to the Anderson model case --- the quantum metric is dominated by optical transitions between hybridizing stretched-exponential topological zero mode resonances. These resonances are typically situated above the crossover energy $\log^2 E_{\rm cut} = 1/\delta^2$ (see~\cite{Note1, komi2} for the discussion of the critical states in this system). Using the expression for the matrix element of the position operator evaluated between two resonances obtained in \cite{komi2} and the Dyson expression for the density of states valid for this model at low energies \cite{eggarter, evers_anderson_2008}
\begin{equation}
|\braket{\psi_E | \hat x | \psi_{- E}}|_{\rm crit} \sim \log^2 |E| \, , ~ \nu_{\rm crit}(E) \sim (|E| \log^3 |E|)^{-1}\, ,
\label{critscale}
\end{equation}
we estimate $g$ using Eq.~\eqref{g} at small $\delta$
\begin{equation}
g \sim \int_{E_{\rm cut}} dE \, \nu_{\rm crit}(E) \, |\braket{\psi_E | \hat x | \psi_{-E}}|_{\rm crit}^2 \sim \delta^{-2} \, .
\label{gcrit}
\end{equation}
This expression matches the well-known critical scaling of the average localization length in this model \cite{evers_anderson_2008, gogolin_singularities_1984, balents_delocalization_1997}, and we immediately obtain $g \sim \xi_{\rm av}$ in the disorder-dominated region $\delta \simeq 0$.

\begin{figure*}[h!tbp]
    \centering
    \includegraphics[width=0.95\textwidth]{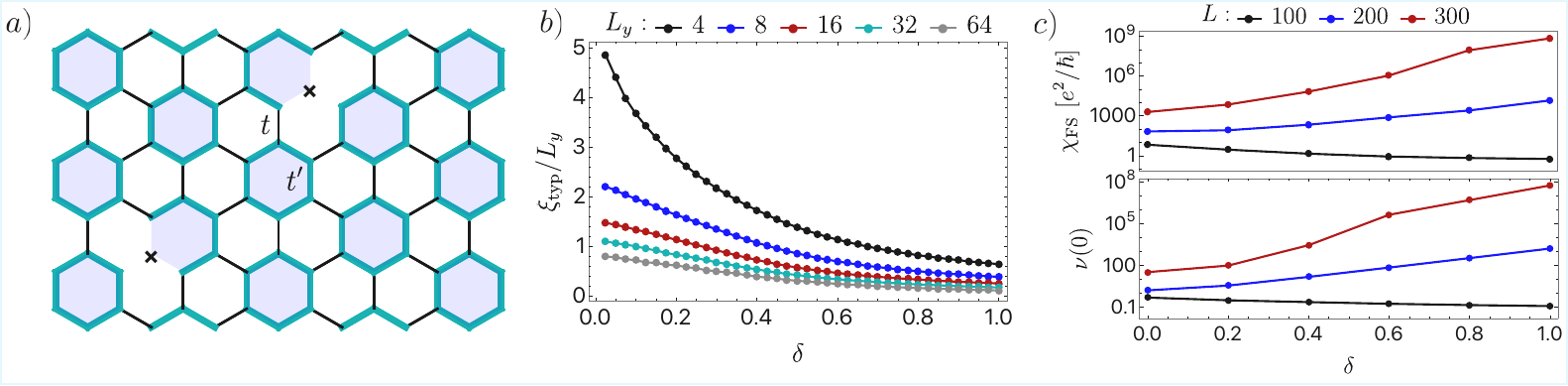}
    \caption{a) A honeycomb lattice with Kekul\'e-O bond order: the black lines correspond to hoppings $t$ while the blue lines indicate hoppings $t' = e^{\delta} t$. Vacancies are introduced by removing random lattice sites with a probability $p = 0.01$ and shown with black crosses. b) The scaling of the ratio of the localization length $\xi_{\rm typ}$ at $E = 0$ to the transverse size of the Kekul\'e graphene ribbon $L_y$ indicates the insulating behavior. c) At the same time, the electric susceptibility and the density of states at $E = 0$ diverge, confirming the emergence of a superdielectric phase. To compute the localization length $\xi_{\rm typ}$, we use the transfer matrix method, while $\chi_{\rm FS}$ is obtained from the small number of eigenstates concentrated around $E = 0$~\cite{Note1}.}
    \label{fig4}
\end{figure*}

We verify these predictions numerically using box disorder with strength $s^2 = 1/5$, as shown in \autoref{fig2}. The simulations confirm that $g$ is proportional to the average (and not the typical localization length, which behaves differently) at small dimerization, while it coincides with the disorder-free quantum metric $g_{\rm clean} \simeq 1/4$ at large $\delta$. Notice that $g_{\rm clean}$ obeys the same $\delta^{-1}$ scaling at criticality as the typical localization length $\xi_{\rm typ}$ since both of these quantities do not receive contributions from the optical transitions between rare hybridized zero modes, responsible for the $g \sim \xi_{\rm av} \sim \delta^{-2}$ scaling.

In summary (see \autoref{tab1}), we find that the scaling relation $g \sim \xi_{\rm av}$ holds near criticality in both the Anderson and SSH chains. This is despite the different nature of the quantum critical regime in these two models, which in the latter assumes a distinctly interband character, driven by the level repulsion between zero modes localized on different sublattices~\cite{Note1}. The proportionality between the quantum metric and the average localization length further implies that the representative spatial size of the interband excitation in both systems is proportional to the geometric average of the lattice constant and the average localization length $\ell_{g} \sim \sqrt{a \cdot \xi_{\rm av}}$. We therefore establish the quantum metric as a direct and universal complementary probe of the average localization length, defined as the decay of the average conductivity~\cite{Note1}. 

\paragraph{Superdielectric phase.---} We have established that the quantum metric in the SSH chain with chiral disorder \eqref{sshham} diverges at $\delta = 0$, and is finite otherwise with the system being an insulator. Within the same model, we now analyze the electric susceptibility given by the linear response expression \cite{grosso, vignale},
\begin{equation}
\chi = \frac{2}{L} \sum_{E_m < E_{\rm F} < E_n} \left \llangle \frac{|\braket{n|\hat x|m}|^2}{E_n - E_m} \right \rrangle \, .
\label{chi}
\end{equation}

At criticality, the susceptibility is more singular than the quantum metric due to the energy denominator (cf. Eq.~\eqref{g}) and therefore also diverges. A less trivial fact is that $\chi$ is also divergent for certain values of $\delta > 0$. Using the expressions for the matrix element of the position operator and the density of states away from criticality \cite{komi2, evers_anderson_2008}, we find
\begin{equation}
|\braket{\psi_E|\hat x|\psi_{-E}}| \sim \log |E| \, , \quad \nu(E) \sim |E|^{-1+2 \delta} \, ,
\label{offcritscale}
\end{equation}
which implies that the Fermi‑surface contribution $\chi_{\rm FS}=\int \! dE\, E^{-1}\nu(E)\,|\braket{\psi_E|\hat x|\psi_{-E}}|^2$ diverges for $\delta\le 1/2$ and is finite otherwise. Thus, the system \eqref{sshham} in this regime exhibits perfect static screening facilitated solely by localized states. This is due to the divergence of $\nu(0)$, which drives the average optical gap $\overline{\Delta}\sim(\xi_{\rm typ}\nu(0))^{-1}$ to zero: interband virtual transitions become effectively gapless, as in a metal, while the dc transport is absent. Numerical data shown in Fig.~\ref{fig3} confirms the divergence of both $\chi_{\rm FS}$ and $\nu(0)$ for $\delta \le 1/2$: a phenomenology analogous to the one observed in certain Griffiths phases in disordered magnets \cite{vojta, sus1, sus2, sus3}.

The superdielectric phase can be realized in a gapped system of any dimensionality, provided that impurity states accumulate strictly at the same energy (suppose, $E = 0$). The low-energy density of states produced by the resonating impurities is then found from the probability of observing a rare region of the volume $V \sim r^D$, in which only two zero modes at the distance $r$ appear  
\begin{equation}
\nu(E) \sim \sum_r e^{- \alpha r} \delta(E - E_0 e^{-r/\xi_{\rm typ}}) \sim E^{-1 + \alpha \xi_{\rm typ}} \, ,
\end{equation}
where $\alpha$ is a model-dependent positive constant with $\alpha \sim |\log (1- p)|$ in the case of vacancies \cite{komi2, grgr1}. At sufficiently large gap (small $\xi_{\rm typ}$) or small impurity concentration $p$, $\nu(E)$ becomes singular, and perfect screening emerges.

A realistic example of such a system in two dimensions is vacancy-doped graphene with bond distortion (see \autoref{fig4}a). The vacancy states accumulate at $E = 0$ \cite{gadevac}, while bond distortion opens a gap, localizing the impurity states, simultaneously preserving the chiral symmetry \cite{kek1, kek2}, which protects the divergence of $\nu(0)$ \cite{grgr1, grgr2}. By diagonalizing this system numerically for different dimerization parameters $\delta$, we observe that the Fermi surface contribution to the electric susceptibility indeed strongly diverges for large dimerization, together with $\nu(0)$, as shown in \autoref{fig4}c, while the system remains localized (see \autoref{fig4}b).

\paragraph{Conclusion.---} 

We studied the relation between three quantities encoding the localization of the system of electrons: the average localization length $\xi_{\rm av}$, the quantum metric $g^{\mu\mu}$, and the electric susceptibility $\chi^{\mu\mu}$. Using the examples of a 1D Anderson model and the SSH chain with bond disorder, we established the scaling $g \sim \xi_{\rm av}$ when localized states are present at the Fermi surface. This relation provides direct access to the average localization length using the set of eigenstates $\ket{n}$, $E_n<E_{\rm F}$, sufficient to compute the quantum metric. By applying the hybridization arguments from our previous study \cite{komi2} to determine $g$, we provided a simple and elegant proof of the $\xi_{\rm av} \sim \delta^{-2}$ critical scaling in the SSH chain with chiral disorder.

We also analyzed the localization properties of disordered chiral insulators and identified a regime characterized by a finite quantum metric and a divergent susceptibility, and therefore, perfect screening of the applied external static electric field. In these materials, states created by certain impurities, such as vacancies, are pinned at a specific energy. However, due to the positional disorder, these states are unable to hybridize into a band and can only couple weakly in a pairwise manner. The resulting electronic correlations are extremely long-range and give rise to Mott resonances with very small energy gaps, resulting in a divergent density of states and enabling perfect screening. Despite their long-range character, the correlations are not sufficiently strong to support finite dc conductivity, and the material remains insulating. We predict that such a superdielectric regime arises in graphene with vacancies and Kekul\'e dimerization, and can be directly accessed via the regular measurement of the dielectric constant through the capacitance or reflectivity.

\begin{acknowledgments}
\emph{Acknowledgments ---} We thank Gil Refael for insightful discussions. This research is supported by the Schwinger Foundation and the National Science Foundation CAREER Award No. DMR-2340394.
T.H.\ acknowledges financial support by the European Research Council (ERC) under grant QuantumCUSP (Grant Agreement No. 101077020). 
\end{acknowledgments}

\bibliography{lit}

\pagebreak
\onecolumngrid
\newpage
\appendix
\begin{center}
\textbf{\large Supplementary information for\\ ``Superdielectrics: Disorder-induced perfect screening in insulators"}
\\[6pt]
Ilia Komissarov, Tobias Holder, Raquel Queiroz
\end{center}

\renewcommand{\thefigure}{\arabic{figure}}
\renewcommand{\figurename}{Supplementary Figure}
\renewcommand{\tablename}{Supplementary Table}
\setcounter{figure}{0}

In this Supplementary Information, we outline the details of the diagrammatic resummation in the 1D Anderson model applied to the calculation of the quantum metric. Next, we show that ac conductivity, electric susceptibility, and the quantum metric can all be obtained from the dynamical structure factor. We also derive the simplified expressions for these quantities valid at low disorder. Next, we review the details of the quantum critical behavior in the SSH chain with chiral disorder, and the effect of Mott resonances on the localization length, quantum metric, and electric susceptibility. Lastly, we outline the details of the transfer matrix calculation for the honeycomb lattice with nearest-neighbor hoppings and Kekul\'e distortion.\\

\textit{Notations and conventions:} Gaussian units $e = \hbar = \epsilon_0 = 1$ are utilized throughout. We use $E$ and $E_{\rm F}$ to denote the reference value of the energy and the Fermi energy, respectively. The quantity $\omega$ bears a meaning of the energy difference, hybridization energy, or the frequency of the applied ac electric field expressed in units of energy, depending on the context. The coordinate vector in three spatial dimensions is denoted with $\bfr$, and $|\bfr| = r$, while in one dimension we use $x$. The definition of the typical localization length used in this work coincides with the one used in \cite{feigelman_dielectric_2018}, i.e., $\xi_{\rm typ} =\lim\limits_{L \to \infty} L/( \llangle \log T(L) \rrangle)$, where $T$ is the dimensionless conductance in the chain of length $L$. The average localization length is defined via the average conductance as $\xi_{\rm av} = \lim\limits_{L \to \infty} L/(  \log \llangle T(L) \rrangle)$. Both $\xi_{\rm typ}$ and $\xi_{\rm av}$ are obtained numerically using the transfer matrix method outlined in \cite{kramer}. The box and the Gauss distributions are defined to have the same variance equal to $W/12$ as shown below:
\begin{equation}
    P_{\rm box}(\varepsilon) = W^{-1} \theta(W/2 - |\varepsilon|)\, , \qquad P_{\rm gauss}(\varepsilon) = \frac{1}{\sqrt{2 \pi} \widetilde W} {\rm exp} \left({-\frac{\varepsilon^2}{2 \widetilde W^2}}\right)\, , \qquad \widetilde W = W/\sqrt{12}\, .
\end{equation}
For the numerical simulations performed for the SSH chain with chiral disorder, we use $P_{\rm box}(\varepsilon)$ as defined above and fix the sum of the inequivalent hoppings to $t+t' = 2$: the remaining free parameters of the model are therefore the difference $t-t'$ and the disorder strength $W$ (equivalently, one can use the logarithmic variables $\delta$ and $s$). The edge states are removed from the spectrum if present. To obtain the quantum metric from the tight-binding model, a choice of the embedding of atoms in the real space is needed: we assume that the length of the hopping $t'$ in the real space is $a=1$, while for $t$ it is zero.

The matrix elements of the position operator in finite systems with periodic boundary conditions, such as the graphene with Kekule dimerization, are computed using the quantity $\hat z = e^{\frac{2 \pi i \hat x}{L}}$ as
\begin{equation}
\braket{n|\hat x|n} \simeq \frac{L}{2 \pi i} \log \braket{n| \hat z|n}\, , \qquad \braket{n|\hat x^2|n} \simeq \left(\frac{L}{2 \pi i}\right)^2 \left[ \log |\braket{n|\hat z|n}|^2 + \log^2 \braket{n|\hat z| n} \right] \, .
\end{equation}
These relations become exact in localized systems in the limit $L \to \infty$, which can be verified directly by expanding the exponents \cite{restax}. For numerical calculations in Kekul\'e graphene, we set the distance between the nearest-neighbor atoms to unity.

\section{Quantum metric from Berezinskii diagrams}
\label{appa}

\begin{table}[t!]
\begin{center}
\begin{tabular}{|c|c|c|c|c|c|c|}
\hline
\multicolumn{7}{|c|}{1D Anderson model} \\
\hline
model & $E_{\rm F}/t$ & disorder & $W/t$ & $\xi_{\rm typ}=\xi_{\rm av}/4$ & $g$ & $g/\xi_{\rm typ}$ \\
\hline
continuum & any & noise & any & --- & --- & $0.1289$ \\
\hline
tight-binding & 0 & box & 0.7 & 107(1) & 14.6 & 0.139(1) \\
\hline
tight-binding & 0.5 & box &  0.7 & 91.3(5) & 12.4 & 0.138(1) \\
\hline
tight-binding & 1 & gauss &  0.7 & 73.5(4) & 10.2 & 0.141(1) \\
\hline
tight-binding & 1.5 & gauss & 0.5 & 84.1(5) & 11.7 & 0.139(1) \\
\hline
\end{tabular}
\end{center}
\caption{Data illustrating the one-parameter scaling of the quantum metric in 1D Anderson model \eqref{andham} ($g \sim \xi_{\rm av}$). For a free electron in a Gaussian noise potential with sufficiently small disorder ($\xi_{\rm typ} \gg 1$), the ratio is $g/\xi_{\rm typ}\simeq 0.1289$. In the full Anderson model \eqref{andham}, this ratio is given by a very similar ratio $g/\xi_{\rm typ} \simeq 0.14$ that does not depend on the disorder strength, disorder type, or position of the Fermi level.}
\label{tabsup}
\end{table}

In this Appendix, we review the diagrammatic method used to obtain the quantum metric in the 1D Anderson insulator. This method utilizes the Dyson equations obtained by Berezinskii \cite{berez}, which incorporate all impurity scattering diagrams for ac conductivity to the zeroth order in the small parameter $(k_{\rm F} \xi_{\rm typ})^{-1}$, e.g., in the limit of low disorder. These relations were formulated initially for the free particle in a Gaussian noise potential; however, we expect that they also apply to low-energy transport properties of any 1D Anderson model falling within the same universality class. Thus, we are interested in the following recursion relations:
\begin{equation}
\begin{gathered}
2 i \pi \xi_{\rm typ} \nu(E_{\rm F}) \omega R_m +m\left(R_{m+1}+R_{m-1}-2 R_m\right)=0 \, , \\
2 i \pi \xi_{\rm typ} \nu(E_{\rm F}) \omega (m+1/2) Q_m+(m+1)^2\left(Q_{m+1}-Q_m\right)-m^2\left(Q_m-Q_{m-1}\right)+R_m-R_{m+1}=0\, ,
\end{gathered}
\end{equation}
with the boundary conditions
\begin{equation}
R_0=1 \, , \qquad  i \pi \xi_{\rm typ} \nu(E_{\rm F}) \omega Q_0+Q_1-Q_0+R_0-R_1=0 \, .
\end{equation}
From the set of complex numbers $\{R_m,Q_m\}$, the ac conductivity is obtained as
\begin{equation}
\sigma(\omega) = \frac{\xi_{\rm typ}}{\pi} {\rm Re}\, \sum_m Q_m\left(R_m-R_{m+1}\right) \, ,
\end{equation}
whereas the electric susceptibility and the quantum metric are subsequently found from the sum rules \cite{swm}
\begin{equation}
\chi = \frac{2}{\pi} \int_0^\inf \frac{d \omega}{\omega^2} \sigma(\omega) \, , \qquad g = \frac{1}{\pi} \int_0^\inf \frac{d \omega}{\omega} \sigma(\omega) \, .
\label{srapp}
\end{equation}

The above procedure is performed using techniques similar to the ones outlined in \cite{gogrec} with the frequency integration cutoff $\omega_0 = 50 (\nu(E_{\rm F}) \xi_{\rm typ})^{-1}$ and $10^7$ integration points. For each frequency $\omega$ in the discretized integration interval $[0,\,\omega_0]$, the $m$-summation cutoff is taken to be $M \simeq 100/\omega$. Due to the fast numerical divergence of $R_m$ and $Q_m$ with $m$, long variables with as many as 200 digits were utilized. The outlined calculation yields $\chi = (4.8\ldots) \nu(E_{\rm F}) \xi_{\rm typ}^2$, fully consistent with \cite{feigelman_dielectric_2018}, and $g = (0.1289\ldots) \xi_{\rm typ}$. Note that the result for the quantum metric is more precise due to this quantity being less susceptible to the disorder fluctuations, mostly affecting the low-frequency contributions in the sum rules \eqref{srapp}.

Expressed in terms of $\ell_g \equiv \sqrt{n^{-1} g} \sim \sqrt{n^{-1} \cdot \xi_{\rm typ}} \sim \sqrt{n^{-1} \cdot \xi_{\rm av}}$ \cite{evers_anderson_2008}, the result for the quantum metric implies that the typical localization size of an electron within the entire $E < E_{\rm F}$ ground state manifold is given by the geometric average between the interparticle distance $n^{-1}$, and the average localization length $\xi_{\rm av}$. The scale $n^{-1}$ appears due to only a small fraction of electrons participating in optical transitions. In order to estimate this fraction, one utilizes the picture of the localization wells of the typical size $\xi_{\rm typ}$ created by disorder (see \autoref{appplot1}a): the typical energy distance between the electrons localized within such wells is $\delta_{\xi} = (\nu(E_{\rm F})\xi_{\rm typ})^{-1}$. Since the electrons are only significantly spatially correlated within the energy window of the order $\delta_{\xi}$ around Fermi, only the fraction
\begin{equation}
\eta = \frac{\nu(E_{\rm F}) \delta_\xi L}{N} = \frac{n^{-1}}{\xi_{\rm typ}}
\end{equation}
of the total number of electrons around the Fermi surface contributes to the dipole fluctuations. Taking into account that the average spread of each electron around the Fermi surface is $\xi_{\rm av} \equiv \xi_{\rm av}(E_{\rm F})$, we estimate the root mean square ground state electron size as
\begin{equation}
\ell_g \sim \sqrt{\eta \cdot \xi_{\rm av}^2} \sim \sqrt{n^{-1} \cdot \xi_{\rm av}} \, ,
\end{equation}
where we used the proportionality $\xi_{\rm av} \sim \xi_{\rm typ}$ \cite{evers_anderson_2008}.

\section{Dynamical structure factor and its relation to transport coefficients and the quantum metric}
\label{appb}

\begin{figure}[h!tbp]
    \centering
    \includegraphics[width=0.99\textwidth]{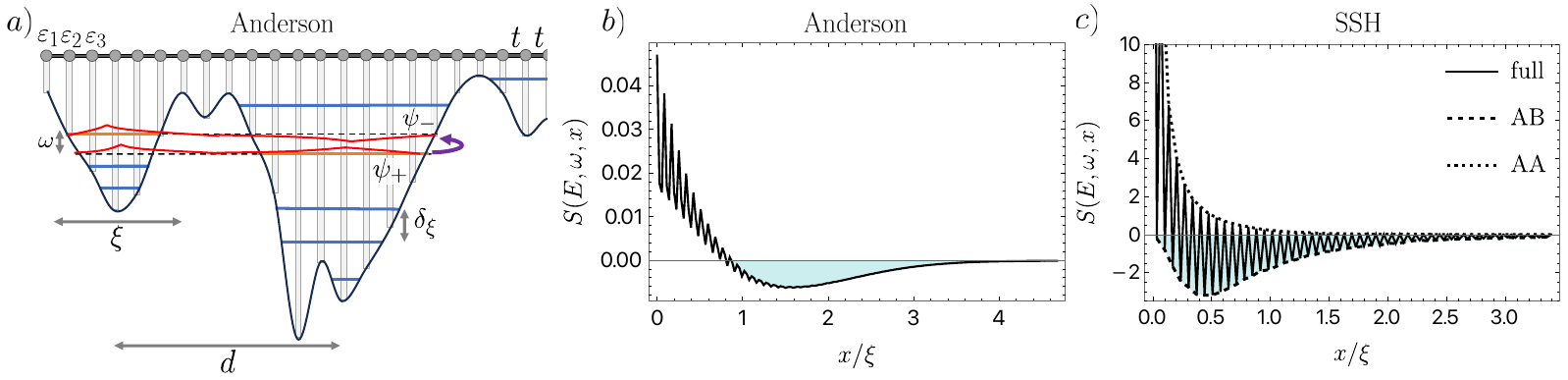}
    \caption{$a)$ A pictorial representation of electronic eigenstates in a 1D Anderson insulator. Electrons, shown with blue lines, are trapped by the disorder within potential wells of the typical size $\xi_{\rm typ}$. The typical energy level spacing within each well is given by the scale $\delta_{\xi} \sim (\nu \xi_{\rm typ})^{-1}$. A subset of electronic eigenstates within different potential wells (shown in orange) match in energy and hybridize pair-wise into extended bonding ($\psi_+$) and antibonding ($\psi_-$) orbitals. Due to the enhancement in length relative to the unhybridized eigenstates, the transition amplitude between such states is also enhanced as $\braket{\psi_-|\hat x|\psi_+} \sim \xi_{\rm typ} \log \omega$ due to the exponential relation between the hybridization energy and the distance $\omega \sim  e^{-d/\xi_{\rm typ}}$. $b)$ Such hybridization events produce negative correlations between electrons and manifest in the appearance of the $S(E,\omega,x)<0$ region in the dynamical structure factor, shaded with blue. Here, box disorder with $W=1$, $E_{\rm F} = 1$, $\omega=0.05$ is used. $c)$ In the SSH chain with chiral disorder (box, $W=5$, $E_{\rm F} = 0$, $t-t'=0.2$, $\omega = 0.005$), the zero modes at different sublattices hybridize away from zero energy, causing the negative level correlations in the AB channel, as showcased in the behavior of the dynamic structure factor. This region gives rise to the topological criticality at $|t|=|t'|$ manifested in the divergence of the quantum metric.}
    \label{appplot1}
\end{figure}

In this Appendix, we discuss how the quantum metric $g$, electric susceptibility $\chi$, and ac conductivity $\sigma(\omega)$ relate to the dynamical structure factor defined as
\begin{equation}
        S(E, \omega, r) =- \frac{1}{\pi^2 \nu(E)}\llangle{\rm Im}\, G^+_{E}\, {\rm Im}\, G^+_{E+\omega}\rrangle= \frac{1}{\nu(E)} \llangle\sum_{mn}\delta(E - E_n) \delta(E +   \omega -E_m) \psi^*_n(\bfr)\psi_m(\bfr)\psi^*_m(0)\psi_n(0) \rrangle \, ,
        \label{scorr}
\end{equation}
where $G^+(E)$ is the retarded Green's function, $\nu(E)$ is the density of states per unit volume, and the double brackets $\llangle \ldots \rrangle$ denote the average over the disorder configurations (we reserve the notation $\braket{...}$ for the matrix element of an operator). The sign convention for the dynamic structure factor is such that the positive values of the function $S(E,\omega,r)$ correspond to the eigenstate attraction, i.e., it is more likely to find two electrons with energies $E$ and $E + \omega$ at the distance $\bfr$ for larger $S$. The negative values of $S$ signify level repulsion arising due to the ability of the localized states to hybridize. The longitudinal ac conductivity, the electric susceptibility, and the quantum metric can all be obtained as certain integrals taken with the correlator $S(E,\omega,r)$, as follows from their linear response expressions \cite{berez, gorkov_structure_nodate}:
\begin{equation}
    \begin{aligned}
        {\rm Re}\, \sigma^{\mu \mu}(\omega) & = - \frac{ \pi   \omega}{2  } \int_{-\inf}^{\inf} dE \,  \nu(E) (f_{E+\omega}-f_{E}) \int d^D\bfr \, (r^\mu)^2 S (E,\omega, r) \, , \\
        \chi^{\mu \mu} & = -\frac{1}{2   } \int_{0}^{\inf} d\omega \int_{- \inf}^{\inf} d E\, \nu(E) \frac{f_{E+\omega} - f_{E}}{\omega} \int d^D\bfr \, (r^\mu)^2 S (E,\omega,r)\, , \\
        g^{\mu \mu} 
        & = - \frac{1}{2 } \int_0^\inf d\omega \int_{-\inf}^{\inf} d E \, \nu(E) (f_{E + \omega} - f_E) \int d^D \bfr\, (r^\mu)^2 S(E,\omega, r) \, .
    \end{aligned}
    \label{schig1}
\end{equation}

We now discuss the behavior of the dynamical structure factor with distance for both the Anderson model and the model of the SSH chain with hopping disorder. In the former model, the properties of $S(E, \omega, r)$ were explained in \cite{gorkov_structure_nodate, gorkov_energy_1983, ivanov_hybridization_2012}. The function $S(E,\omega,x)$, shown in \autoref{appplot1}$a$, takes non-zero values in two coordinate regions. In the region $0<x \lesssim \xi_{\rm typ}$, $S(E,\omega,x)$ is positive due to the trapping of multiple electrons in the localization volumes of the typical size $\xi_{\rm typ}$ leading to level attraction. The second coordinate region is associated with a much larger length scale referred to as the Mott length $d(\omega) \sim \xi_{\rm typ} \log(\omega)$ \footnote{The hybridization energy split is exponentially small in the distance between the electrons $\omega \sim \delta_{\xi} e^{-2 d/\xi_{\rm typ}}$. Solving this for $d$ gives the Mott scale $d(\omega)$.} and corresponds to the rare resonances appearing when two electrons belonging to different localization volumes match in energy.
 
In the case of the SSH chain with chiral disorder, shown in \autoref{appplot1}$b$, a similar analysis applies. A crucial difference, however, is that the level repulsion in this case is only present between the states localized at different sublattices. These interband correlations arising from the hybridization of the topological zero modes give rise to the quantum critical glassy behavior analyzed in detail in \cite{komi2}.

\section{Simplified expressions for the transport coefficients near criticality}

\label{appc}

In this Appendix, we show that when ($\xi_{\rm typ} \gg 1$), the following expressions hold true:
\begin{equation}
    \begin{aligned}
        {\rm Re}\, \sigma_{\rm FS}^{\mu\mu}(\omega) & \simeq - \frac{\pi \omega^2 \nu(E_{\rm F})}{2   } \int d^D \bfr \, (r^\mu)^2 S(E_{\rm F},\omega, r) \, , \\
        \chi_{\rm FS}^{\mu \mu} & \simeq  \frac{1}{2  } \int d^D \bfr \, (r^\mu)^2 \left \llangle \sum_n \delta(E_{\rm F} - E_n) |\psi_n(\bfr)|^2 |\psi_n(0)|^2 \right \rrangle \, , \\
        g_{\rm crit}^{\mu \mu} & \simeq - \frac{1}{2 } \int d^D \bfr \, (r^\mu)^2 \left \llangle \sum_{m > n} \delta(E_{\rm F} - E_n) (E_m - E_n) \psi^*_n(\bfr)\psi_m(\bfr)\psi^*_m(0)\psi_n(0) \right \rrangle \, .
    \end{aligned}
    \label{trapprox}
\end{equation}
The first two relations are known (see, e.g. \cite{lgp, berdenden, feigelman_dielectric_2018}), while the one for the quantum metric is new. It shows that at low disorder, only the transitions between the states at Fermi energy to the ones above $E_{\rm F}$ are relevant, reducing the number of terms in the sums over eigenstates from $O(N^2)$ to $O(N)$. This is in stark contrast with $\chi_{\rm FS}$, which simplifies to the sum over eigenstates close to the $E_{\rm F}$, and therefore represents a Fermi surface property when $\xi_{\rm typ} \gg 1$. The expressions 1 and 3 follow from 1 and 3 in \eqref{schig1} by expanding the Fermi functions at low frequency $f_{E + \omega} - f_{E} \simeq - \omega \delta(E - E_{\rm F})$ \footnote{Another way to reproduce the same expression is by assuming that the dynamical structure factor $S(E, \omega, r)$ is $E$-independent, as e.g. happens in the Anderson model at low disorder.}, while the relation for $\chi$ requires some additional work \cite{berdenden}, which we reproduce below.

Firstly, using the orthogonality of states and the homogeneity property of the disorder average, one obtains the normalization conditions
\begin{equation}
    \int d \omega\, S(E, \omega, r) = \delta(\bfr)\, , \qquad \int d^d \bfr\, S(E, \omega, r) = \delta(\omega) \, .
    \label{appnorms}
\end{equation}
Following \cite{berdenden}, in the definition of the dynamic structure factor \eqref{scorr}, we concentrate on the $m = n$ terms. Out of these terms, only the ones corresponding to the localized states contribute to $S(E, \omega, r)$: otherwise, the contribution vanishes as $1/V$ in the thermodynamic limit. Therefore, if the states in the vicinity of the energy $E$ are localized, we can expand
\begin{equation}
    S(E,\omega,r) = \delta(\omega) F(E,r) + G(E,\omega, r) \, ,
    \label{sfg}
\end{equation}
where
\begin{equation}
    F(E,r) = \frac{1}{\nu(E)}\left\llangle\sum_n \delta(E_n-E)\left|\psi_n(\bfr)\right|^2\left|\psi_n(0)\right|^2\right\rrangle \, .
    \label{frel}
\end{equation}
Using the first normalization condition in \eqref{appnorms}, we find
\begin{equation}
    \int d\omega\, G(E,\omega, r) = - F(E,r) \, , ~ \bfr \neq 0 \, .
    \label{gfrel}
\end{equation}
Upon expanding the Fermi functions in the second expression in \eqref{schig1} in $\omega$ and subsequently using \eqref{sfg}, \eqref{frel}, \eqref{gfrel}, the second line in \eqref{trapprox} follows.  

We now discuss the applications of \eqref{schig1} in the case of a 1D Anderson insulator, focusing first on low-frequency ac conductivity. At small $\omega$, the integral for $\sigma(\omega)$ is dominated by the Mott pair minimum in the dynamical structure factor (see \autoref{appplot1} $a, \, b$). The behavior of $S(E, \omega, x)$ in the vicinity of this minimum can be cast as \cite{gorkov_structure_nodate} 
\begin{equation}
    S(E, \omega, x) \simeq -\nu \left(E\right)\left(\frac{\xi_{\rm typ}}{4 \pi d(\omega)}\right)^{1 / 2} \exp \left(-\frac{\left[x-d(\omega)\right]^2}{4 \xi_{\rm typ} d(\omega)}\right) \, .
\end{equation}
Performing the position integral in the expression \eqref{schig1} for $\sigma(\omega)$ then yields the Mott-Berezinskii law \cite{berez, abrikosov_conductivity_1978}
\begin{equation}
    {\rm Re}\, \sigma(\omega) =  4 \pi \xi_{\rm typ}  \left(\frac{\omega}{\delta_{\xi}}\right)^2 \ln ^{2} \frac{{\bar \delta}_{\xi}}{\omega}\, ,
    \label{mbapp}
\end{equation}
where ${\bar \delta}_{\xi} = (\pi \nu \xi_{\rm typ})^{-1}$. The $\log^2 \omega$ factor in this expression results from the optical transitions between Mott resonances (see \autoref{appplot1}$a$). One also anticipates that Mott resonances, being the most spatially extended states, should provide the dominant contribution to the electric susceptibility and the quantum metric, estimated as \cite{swm}
\begin{equation}
\chi \simeq \frac{2}{\pi} \int_0^{\bar \delta_{\xi}} \frac{d\omega}{\omega^2} \, {\rm Re}\, \sigma(\omega) \simeq 5.1\, \nu(E_F) \xi_{\rm typ}^2\, , \qquad g \simeq \frac{1}{\pi} \int_0^{\bar\delta_\xi} \frac{d \omega}{\omega} \, {\rm Re} \, \sigma(\omega) \simeq 0.1 \,\xi_{\rm typ}.
\label{chigapp}  
\end{equation}
The above estimates agree well with $\chi_{\rm} = (4.808\ldots) \cdot \nu(E_F) \xi_{\rm typ}^2$ found in \cite{feigelman_dielectric_2018} and $g = (0.1289\ldots) \cdot \xi_{\rm typ}$ obtained in the main text: therefore, both quantities are indeed dominated by the optical transitions between Mott resonances.

Importantly, it follows from \eqref{mbapp}, \eqref{chigapp} that the optical conductivity, dielectric constant, and the quantum metric in Anderson insulators do not depend on the details of the disorder distribution and the position of the Fermi energy (see also the data in \autoref{tabsup}). This is a consequence of the universality of the dynamic structure factor in the Anderson model. The non-universal Friedel oscillations seen in \autoref{appplot1}$a,\,b$ become fast in comparison with the smooth spatial profile of $S(E, \omega, x)$ varying on a scale $\xi_{\rm typ}$ at low disorder, and therefore do not contribute to the position integrals in \eqref{trapprox}.\\

\section{Topological criticality and zero-mode resonances in the SSH chain}

\label{appe}

\begin{figure}[t!tbp]
    \centering
    \includegraphics[width=0.9\textwidth]{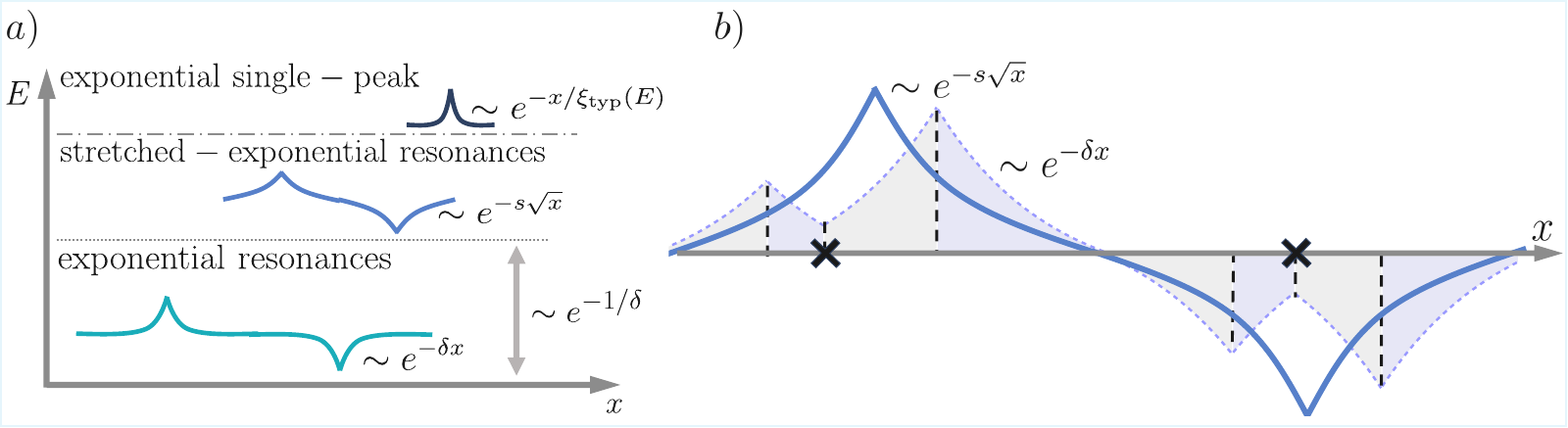}
    \caption{$a)$ The scheme showcasing the typical energy ranges corresponding to the different types of states present in the SSH chain with chiral disorder \eqref{sshapp}, reprinted from~\cite{komi2}. Away from criticality ($\delta > 0$), in the vicinity of $E = 0$ lie the resonances formed by the exponentially localized zero modes. Above them exist the critical stretched-exponentially localized resonances, which are responsible for the emergence of the critical delocalization at $\delta = 0$. At even higher energies lie the regular, exponentially-localized states not originating from the hybridizing topological zero modes. $b)$ The typical profile of the stretched-exponentially localized state is shown with the thick blue line, while the dashed blue line shows a representative example of such a state. The spatial structure of a stretched-exponentially localized resonance before averaging is represented by the regions with the alternating topological regions, within which the zero mode either decays or grows exponentially: shaded in blue or gray. Typically (see the thick blue envelope), the decay of an individual zero mode, shown with a black cross, is stretched-exponential due to the random walk created by the regions with different $w$. When two such stretched-exponential zero modes hybridize, they produce bonding and anti-bonding (shown) states of the average length $d(E) \sim \xi_{\rm av} \sim \log^2 E$, due to the modified Mott relation between energies and distances $E \sim e^{-\sqrt{d(E)}}$.}
    \label{appsch1}
\end{figure}

In this Appendix, we review the behavior of the typical and the average localization lengths around criticality in the SSH model with chiral disorder. Furthermore, we discuss the properties and the transport manifestations of the Mott resonances formed by hybridizing topological zero modes.

We consider the Su-Schriffer-Heeger model with hopping disorder described by the Hamiltonian
\begin{equation}
    \hat H = \sum_i (t' - \varepsilon_i')\ket{i,\rm B} \bra{i,\rm A}  + (t - \varepsilon_i) \ket{i+1,\rm B} \bra{i,\rm A} + {\rm h.c.}\, ,
    \label{sshapp}
\end{equation}
where $\varepsilon_i$, $\varepsilon_i'$ are drawn from the same disorder distribution, and $t$, $t'$ are arbitrary hoppings in the clean system. This type of disorder preserves the chiral symmetry that acts on the states as
\begin{equation}
    \ket{i,\rm A} \to \ket{i,\rm A} \, , \qquad \ket{i,\rm B} \to -\ket{i,\rm B} \, .
\end{equation}
Under this transformation, each eigenvalue of the Hamiltonian in \eqref{sshapp} at the energy $E$ maps to the one at the energy $-E$. The exception is only the $E=0$ state: this energy is special, and turns out to be the one at which the delocalization transition occurs in the compensated case $|t|=|t'|$. One can establish the form of the divergence of the typical localization length using the Thouless relation for the typical localization length \cite{Thouless1972}
\begin{equation}
    \xi_{\rm typ}^{-1}(E) = \int_{- \infty}^{+ \infty} d E' \, \nu(E') \log|E - E'| - \frac{1}{2} \llangle \log |t' - \varepsilon'||t-\varepsilon| \rrangle\, ,
    \label{thoulapp}
\end{equation}
as shown in \cite{theodorou_extended_1976}. We reproduce this result below.  

At criticality and $E=0$, we simply have
\begin{equation}
    \frac{1}{2} \llangle \log |t' - \varepsilon'||t - \varepsilon| \rrangle = \int_{- \infty}^{+ \infty} d E' \, \nu(E') \log|E - E'| \, .
    \label{thoulapp0}
\end{equation}
Combining \eqref{thoulapp} and \eqref{thoulapp0}, we write for $E \neq 0$
\begin{equation}
    \xi_{\rm typ}^{-1}(E) = 2 \int_0^E dE' \, \nu(E')\log|E-E'| + 2 \int_E^\inf dE' \, \nu(E')\log|E-E'| - 2 \int_0^\inf dE'\, \nu(E') \log |E'| \, .
\end{equation}
For small $E$, we can neglect the $E'$ dependence in the $\log$ of the first integral, and the $E$ dependence in the $\log$ of the second, which leads to
\begin{equation}
\begin{aligned}
    \xi_{\rm typ}^{-1}(E) \simeq\, & 2 \int_0^E dE' \, \nu(E') (\log|E| - \log|E'|) =2 \int_{-E}^E d E'' \int_{E''}^E dE' \frac{\nu(E'')}{E'} \\ = \, & 2\int_0^E dE' \int_{-E'}^{E'} dE'' \frac{\nu(E'')}{E'} =2 \int_0^E \frac{N(E')}{E'} \, ,
\end{aligned}
\label{xin}
\end{equation}
where 
\begin{equation}
    N(E) = \int_{-E}^E d E'' \, \nu(E'') \sim \frac{s^2}{|\log E|^2} 
\label{nofe}
\end{equation}
is obtained from the Dyson singularity describing the low-energy density of states in this model \cite{Krishna_2021}, with
\begin{equation}
    s^2 = \left \llangle \log^2 \left| \frac{t' - \varepsilon}{t}\right| \right  \rrangle - \left \llangle \log \left| \frac{t' - \varepsilon}{t}\right| \right \rrangle^2 
\end{equation}
representing the disorder strength on the logarithmic scale. Plugging \eqref{nofe} into \eqref{xin} and evaluating the integral, we obtain the asymptote of the localization length at low $E$
\begin{equation}
    \xi_{\rm typ}(E) \simeq s^{-2} \log |E| \, .
    \label{xityp}
\end{equation}

\begin{figure}[t!tbp]
        \centering
        \includegraphics[width=0.95\textwidth]{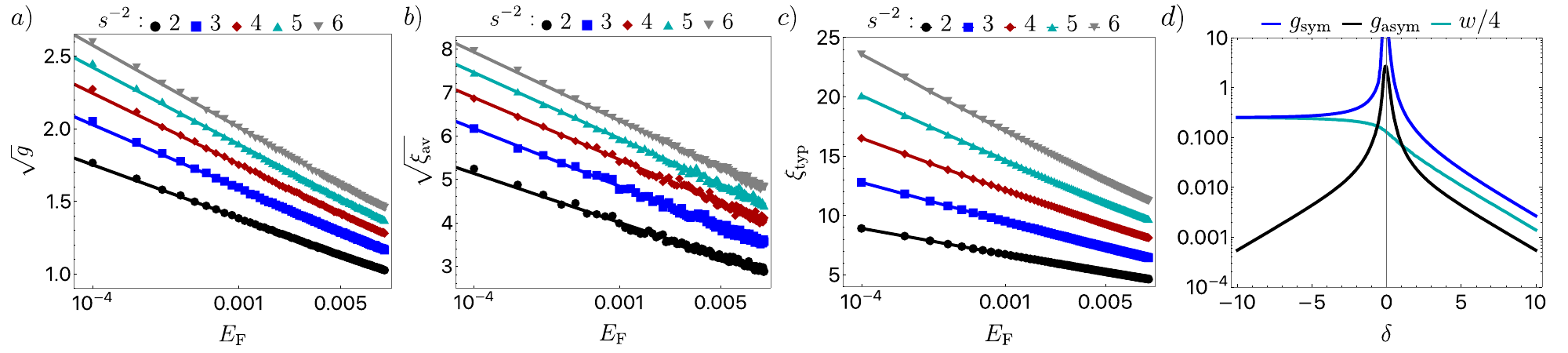}
        \caption{$a)$ quantum metric in the SSH chain with chiral disorder calculated for five different disorder strengths $s^2$ diverges as $\log^2 E_{\rm F}$ at criticality, in agreement with \eqref{gc}. $b)$ Behavior of the average localization length 
       $\xi_{\rm av}$ follows that of the quantum metric at criticality, i.e. $g \sim \xi_{\rm av}$. Despite the fact that we used a comparable number of replicas ($\sim 10^4$) to compute $g$, and $\xi_{\rm av}$, the latter quantity fluctuates stronger; therefore, computing quantum metric provides a better point of access to $\xi_{\rm av}$ than calculating the average conductivity using the transfer matrix. $c)$ The typical localization length diverges as $\log E_{\rm F}$, in agreement with \eqref{xityp}. Unlike the typical localization length, this quantity is obtained from the average of the logarithm of the conductance, which is a self-averaging quantity.  This makes the disorder fluctuations in $\xi_{\rm typ}$ much milder than in $\xi_{\rm av}$. $d)$ Symmetric in energy optical transitions created by the hybridizing zero modes produce a divergent contribution to the quantum metric at criticality $g_{\rm sym}$, while the contribution from the asymmetric transitions $g_{\rm asym}$ does not diverge (as confirmed by reproducing this data for different system sizes). The real-space winding number $w$ \cite{windingr} is also calculated, and the topological bound $g > w/4$ is confirmed. Disorder strength $s^{-2} = 5$ is utilized.}
        \label{appplot2}
\end{figure}

The expression \eqref{xityp} is verified using the transfer matrix method as shown in \autoref{appplot2}$c$. It is important to point out that the localization length above is a \textit{harmonic} ($\xi_{\rm typ}=\llangle \xi^{-1} \rrangle^{-1}$) average, and hence describes the \textit{typical} and not the \textit{average} value of the localization length. The latter turns out to be significantly more singular, and given by the expression \cite{gogolin_singularities_1984, balents_delocalization_1997}
\begin{equation}
    \xi_{\rm av}(E) \sim \log^2 |E|\, .
    \label{xiav}
\end{equation}  
The average localization length can be obtained using the transfer matrix method as a quantity that characterizes the spatial decay of the average conductance $T$ (the typical localization length, on the other hand, characterizes the decay of the $\llangle \log T \rrangle$)
\begin{equation}
\xi_{\rm av} = \lim\limits_{L \to \infty} \frac{L}{ \log \llangle T(L)\rrangle}\, .
\end{equation}
The expression \eqref{xiav} is verified numerically using the expression above in \autoref{appplot2}$b$.

As argued in the main text, the quantum metric is proportional to the localization length near criticality in 1D. In the case of the SSH chain with chiral disorder, the dominant contributions to the quantum metric arise from the symmetric in energy transitions between stretched-exponentially localized zero modes \cite{komi2}
\begin{equation}
g(E_{\rm F}) \simeq \int_{\rm E_F} \nu(E) d E \, |\braket{\psi_E|\hat x| \psi_{-E}}|^2 \sim \log^2 E_{\rm F}\, ,
\label{gc}
\end{equation}
where we used the Dyson expression for the density of states $\nu(E) \sim (|E| \log^3 E)^{-1}$ and the behavior of the matrix element of the position operator taken between the bonding and antibonding resonances formed by two stretched-exponentially localized impurity states $|\braket{\psi_E|\hat x| \psi_{-E}}| \sim \log^2 E$. As anticipated, the critical behavior in \eqref{gc}, verified numerically in \autoref{appplot2}$a$, matches that of the average localization length \eqref{xiav}, shown in \autoref{appplot2}$b$. Analogously, one finds the leading in $E_{\rm F}$ behavior of the electric susceptibility 
\begin{equation}
\chi(E_{\rm F}) \simeq \int_{\rm E_F} \frac{\nu(E) d E}{E} |\braket{\psi_E|\hat x| \psi_{-E}}|^2 \sim \frac{\log E_{\rm F}}{E_{\rm F}}\, .
\label{xicrapp}
\end{equation}

The special role played by the symmetric around $E_{\rm F} = 0$ transitions at criticality can be illustrated by considering the behavior of the quantum metric, which we expand as $g = g_{\rm sym} + g_{\rm asym}$ as a function of the hopping staggering parameter $\delta$ shown in \autoref{appplot2}$d$. While the contribution due to the asymmetric in energy interband transitions $g_{\rm asym}$ is always finite, $g_{\rm sym}$, arising due to the transitions between the states at the energies $-E$ and $E$, diverges at the critical value $\delta = 0$. The asymmetric contribution due to the low-energy states can be cast as
\begin{equation}
    g_{\rm asym} = \int^0_{-\infty} dE\, \nu(E) \int_0^\infty dE'\, \nu(E') \left.|\braket{\psi_E|\hat x|\psi_{E'}}|^2\right._{E \neq E'} \, .
\end{equation}
Since the low-energy density of states at criticality is given by the Dyson singularity $\nu(E) \sim (|E| 
\log^3 |E|)^{-1}$, and the matrix element of the position operator is regular for asymmetric transitions, due to the symmetric relative to $E_{\rm F} = 0$ pattern of hybridization of the zero modes, the logarithmic factors in the density of states make the integrals converge. Therefore, $g_{\rm asym}$ is finite in consistency with the numerical data, and the divergence of $g$ is caused entirely by the symmetric in energy transitions (see \eqref{gc}).

\section{Transfer matrix in the Kekul\'e-distorted graphene}

\label{appf}

\begin{figure}[t!]
        \centering
        \includegraphics[width=0.32\textwidth]{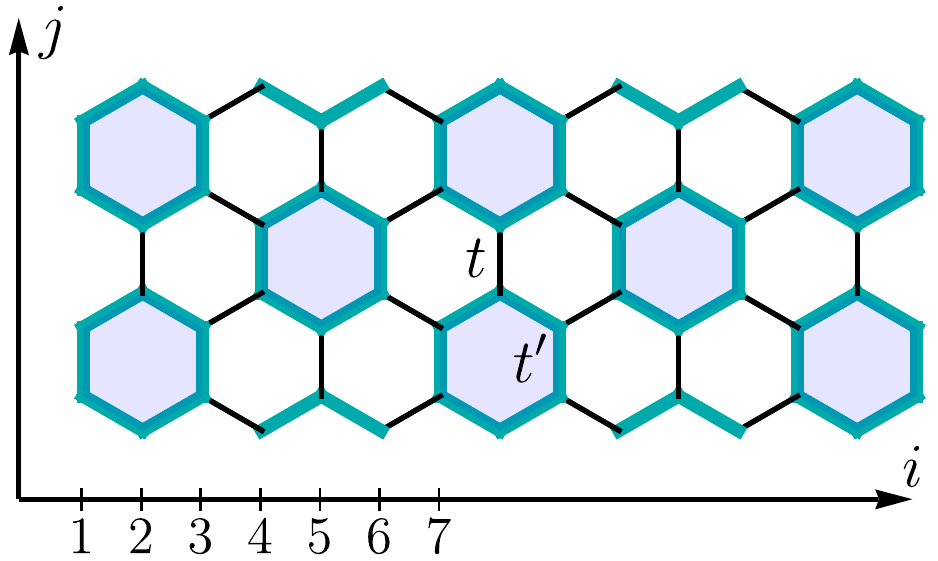}
        \caption{Schematics of the graphene nanoribbon with Kekul\'e-O distortion ($L_x = 15$, $L_y = 4$) utilized for the transfer matrix calculations.}
        \label{appplot3}
\end{figure}

In this Appendix, we provide the details of the transfer matrix calculation of the localization length in Kekul\'e-O distorted graphene. The figure \eqref{appplot3} shows the utilized setup. Periodic boundary conditions along the boundaries in the propagation direction are used in order to eliminate the influence of boundary modes.

We label the values of the wavefunction in the real space using the index $i$ growing in the direction of propagation, as shown in \autoref{appplot3}, and index $j$ growing in the transverse direction. The transfer matrix relates two sets of wavefunctions: $\{ \psi_{i-2,1}, \ldots \psi_{i-2,M}, \psi_{i-1,1}, \ldots \psi_{i-1,M} \}$ to the analogous set where the $i$ coordinate is shifted by 1, i.e. $\{ \psi_{i-1,1}, \ldots \psi_{i-1,M}, \psi_{i,1}, \ldots \psi_{i,M} \}$. The transfer matrix needed to advance from $i=1$ to $i=2$  is different from the one needed to shift from $i=2$ to $i=3$. The chain of multiplications only repeats when one propagates from $ i = 1$ to $i=7$. Thus, by solving the Schr\"odinger equations written on layers $i=2$ to $i=6$, one finds the minimal set of six transfer matrices that can be parameterized as
\begin{equation}
{\cal M} = \begin{pmatrix}
{\cal M}^{11} & {\cal M}^{12} \\
{\cal M}^{21} & {\cal M}^{22}
\end{pmatrix} \, .
\end{equation}
It is convenient to introduce $M \times M$-sized matrices: ${\cal I}$ is the identity matrix, ${\cal O}$ is the zero matrix, and two more
\begin{equation}
{\cal M_1}_1 = 
\begin{pmatrix}
0 & 1 & 0 & 0 & 0 & \cdots \\
1 & 0 & 0 & 0 & 0 & \cdots \\
0 & 0 & 0 & 1 & 0 & \cdots \\
0 & 0 & 1 & 0 & 0 & \cdots \\
0 & 0 & 0 & 0 & 0 & \cdots \\
\vdots & \vdots & \vdots & \vdots & \vdots & \ddots
\end{pmatrix} \, , \qquad 
{\cal M_2}_2 = \begin{pmatrix}
0 & 0 & 0 & 0 & 0 & \cdots \\
0 & 0 & 1 & 0 & 0 & \cdots \\
0 & 1 & 0 & 0 & 0 & \cdots \\
0 & 0 & 0 & 0 & 1 & \cdots \\
0 & 0 & 0 & 1 & 0 & \cdots \\
\vdots & \vdots & \vdots & \vdots & \vdots & \ddots
\end{pmatrix} \, .
\end{equation}
Using the above expressions, one can write the components of the six required transfer matrices as
\begin{equation}
\begin{aligned}
\begin{array}{llll}
{\cal M}_{\rm 1 \to 2}^{11} = \dfrac{E}{t'} {\cal I} - {\cal M}_1 \, , &
{\cal M}_{\rm 1 \to 2}^{12} = -\dfrac{t}{t'} {\cal I} \, , &
{\cal M}_{\rm 1 \to 2}^{21} = {\cal I} \, , &
{\cal M}_{\rm 1 \to 2}^{22} = {\cal O} \, , \\[6pt]
{\cal M}_{\rm 2 \to 3}^{11} = \dfrac{E}{t'} {\cal I} - \dfrac{t}{t'} {\cal M}_2 \, , &
{\cal M}_{\rm 2 \to 3}^{12} = -{\cal I} \, , &
{\cal M}_{\rm 2 \to 3}^{21} = {\cal I} \, , &
{\cal M}_{\rm 2 \to 3}^{22} = {\cal O} \, , \\[6pt]
{\cal M}_{\rm 3 \to 4}^{11} = \dfrac{E}{t} {\cal I} - \dfrac{t'}{t} {\cal M}_1 \, , &
{\cal M}_{\rm 3 \to 4}^{12} = -\dfrac{t'}{t} {\cal I} \, , &
{\cal M}_{\rm 3 \to 4}^{21} = {\cal I} \, , &
{\cal M}_{\rm 3 \to 4}^{22} = {\cal O} \, , \\[6pt]
{\cal M}_{\rm 4 \to 5}^{11} = \dfrac{E}{t'} {\cal I} - {\cal M}_2 \, , &
{\cal M}_{\rm 4 \to 5}^{12} = -\dfrac{t}{t'} {\cal I} \, , &
{\cal M}_{\rm 4 \to 5}^{21} = {\cal I} \, , &
{\cal M}_{\rm 4 \to 5}^{22} = {\cal O} \, , \\[6pt]
{\cal M}_{\rm 5 \to 6}^{11} = \dfrac{E}{t'} {\cal I} - \dfrac{t}{t'} {\cal M}_1 \, , &
{\cal M}_{\rm 5 \to 6}^{12} = -{\cal I} \, , &
{\cal M}_{\rm 5 \to 6}^{21} = {\cal I} \, , &
{\cal M}_{\rm 5 \to 6}^{22} = {\cal O} \, , \\[6pt]
{\cal M}_{\rm 6 \to 7}^{11} = \dfrac{E}{t} {\cal I} - \dfrac{t'}{t} {\cal M}_2 \, , &
{\cal M}_{\rm 6 \to 7}^{12} = -\dfrac{t'}{t} {\cal I} \, , &
{\cal M}_{\rm 6 \to 7}^{21} = {\cal I} \, , &
{\cal M}_{\rm 6 \to 7}^{22} = {\cal O} \, ,
\end{array}
\end{aligned}
\end{equation}
To account for the presence of vacancies with concentration $p \%$, in each matrix element containing $E$, a replacement $E-V$ with a probability $p/100$ is made, where $V$ greatly exceeds all the other parameters in the model.

\end{document}